\documentclass[aps,pre,twocolumn,superscriptaddress,showpacs]{revtex4}
\usepackage{times}
\usepackage{color}
\usepackage{float}
\usepackage{epsfig}

\begin{document}

\title{Diversity of chimera-like patterns from a model of 2D arrays of neurons
with nonlocal coupling}

\author{Changhai Tian}
\affiliation{Department of Physics, East China Normal University,
Shanghai 200062, P. R. China}
\affiliation{School of Data Science, Tongren University, Tongren 554300, P. R. China}

\author{Xiyun Zhang}
\affiliation{Department of Physics, East China Normal University,
Shanghai 200062, P. R. China}

\author{Zhenhua Wang}
\affiliation{Department of Physics, East China Normal University,
Shanghai 200062, P. R. China}

\author{Zonghua Liu}
\email{zhliu@phy.ecnu.edu.cn} \affiliation{Department of Physics,
East China Normal University, Shanghai 200062, P. R. China}

\begin{abstract}

Chimera states have been studied in 1D arrays, and a variety of different chimera states have been found using
different models. Research has recently been extended to 2D arrays but only  to phase models of them. Here, we extend it to a nonphase model of 2D arrays of neurons and focus on the influence of nonlocal coupling. Using extensive numerical simulations, we find, surprisingly, that this system can show most types of previously observed chimera states, in contrast to previous models, where only one or a few types of chimera states can be
observed in each model. We also find that this model can show some special chimera-like patterns such as gridding and multicolumn patterns, which were previously observed only in phase models. Further, we
present an effective approach, i.e., removing some of the coupling links, to generate heterogeneous coupling,
which results in diverse chimera-like patterns and even induces transformations from
one chimera-like pattern to another.

\end{abstract}


\maketitle

\section{Introduction}
It is generally believed that the collective behavior of coupled identical oscillators will be either
synchronized or unsynchronized for local or global coupling. However,
in 2002, Kuramoto and his colleagues noticed that it is also possible to observe peculiar patterns in which
the oscillators separate sharply into two groups, one composed of mutually synchronized oscillators with
a unique frequency and the other composed of desynchronized oscillators with distributed frequencies, provided
that nonlocal coupling is adopted \cite{Kuramoto:2002}. The surprising aspect of this phenomenon is that
it was detected in systems of identical oscillators coupled in a symmetric ring topology with a symmetric
interaction function, and it coexists with a completely synchronized state. This highly counterintuitive
phenomenon was then given the name chimera state (CS) in 2004 by Abrams and Strogatz \cite{Abrams:2004}.
Since then, CSs have been the focus of extensive research in a wide number of models, including
neuronal systems \cite{Omelchenko:2013,Hizanidis:2014,Sakaguchi:2006,Olmi:2010}, chaotic oscillators
\cite{Omelchenko:2011,Omelchenko:2012}, high-dimensional systems
\cite{Omel'chenko:2012,Panaggio:2013,Panaggio:2015,Xie:2015,Maistrenko:2015}, and even experimental systems
\cite{Hagerstrom:2012,Tinsley:2012,Viktorov:2014}. To date, CSs can be considered as a variety of self-organized
spatiotemporal patterns in which regions of coherence and incoherence coexist. These patterns include the
multicluster CS \cite{Omelchenko:2013,Yao:2015}, breathing CS \cite{Abrams:2008,Ma:2010}, spiral wave
CS \cite{Panaggio:2013,Xie:2015,Maistrenko:2015,Martens:2010,Gu:2013,Laing:2009}, circular spot CS and
stripe CS \cite{Panaggio:2013,Panaggio:2015}, traveling CS \cite{Xie:2014}, and chimera death
\cite{Zakharova:2014,Dutta:2015,Banerjee:2015}. CS is applied in many realistic systems such as the unihemispheric sleep observed in birds, lizards, and dolphins
\cite{Ma:2010,Panaggio:2015b,Rattenborg:2000}; electrical spiral and scroll waves in the context of the heart
\cite{Cherry:2008}; and epileptic seizures \cite{Rothkegel:2014}.

In light of the above results, an important question becomes, ``What is the next key topic in the
study of CSs?'' To identify this topic, we carefully recheck the above results and find three common
points: (1) Most of the studies are focused on 1D arrays and consider both phase models and nonphase models,
such as the complex Ginzburg--Landau equation and neurons. (2) A few studies have extended the
phenomenon of the CS to 2D arrays, but only in phase models such as Kuramoto oscillators. Further, (3)
each model can show only one or a few types of CS patterns. From these three features, two interesting
questions arise. The first is whether it is possible to extend the CS to 2D systems
of nonphase models, and the second is whether it is possible to use only one system to show most
of the observed CS patterns. In fact, the second question is especially important, as the synchronized or
partially synchronized patterns in neural systems are closely related to the functions of the brain, such as
cognitive and memory functions
\cite{Singer:1999,Hipp:2011,Roelfsema:1997,Vogels:2005,Diesmann:1999,Womelsdorf:2007,Xu:2013,Liu:2012,Wang:2013,Xu:2014}.
If we can provide evidence proving that it is possible for one neural system to show most of the CS
patterns observed to date, it will be extremely helpful for understanding the large capacity of human memory
patterns, which has been understood only in terms of artificial neural networks \cite{Hopfield:1982}.
It is, therefore, of fundamental importance to construct a model to implement the various
CS or similar patterns, collectively referred to as chimera-like patterns, which is the aim of this work.
We here propose such a model of
FitzHugh--Nagumo (FHN) neurons to show that the CS can indeed be extended to 2D arrays of nonphase models and that the
models do show diverse chimera-like patterns. We also find that this model can show some
special chimera-like patterns such as gridding and multicolumn patterns, which were previously
observed only in phase models. Further, we present an effective approach to generating heterogeneous coupling,
i.e., removing some of the coupling links. We find, surprisingly, that heterogeneous coupling can result in
diverse chimera-like patterns and, in particular, can cause transformations from one type of
chimera-like pattern to another.

The paper is organized as follows. In Sec. II, we present the model of 2D arrays of FHN neurons with nonlocal
coupling and introduce the local order parameter to measure their collective behavior. In Sec. III, we present
two methods of nonlocal coupling to generate diverse local synchronized patterns and the new patterns
in our model. In Sec. IV, we discuss the observed patterns and present our conclusions.

\section{Model description}
We consider a model of 2D arrays of $N\times N$ nonlocally coupled, identical FHN neurons with a periodic
boundary condition (the torus configuration). The system is mathematically described as
\begin{eqnarray}
\label{model}
\epsilon\dot{u}_{ij} &=& u_{ij}-\frac{u^3_{ij}}{3}-v_{ij}+\frac{c}{B_r(i,j)} \nonumber \\
& &\sum_{(k,l)\in B_r(i,j)}
[b_{uu}(u_{kl}-u_{ij})+b_{uv}(v_{kl}-v_{ij})],  \nonumber \\
\dot{v}_{ij} &=& u_{ij}+a+\frac{c}{B_r(i,j)} \nonumber \\
& &\sum_{(k,l)\in B_r(i,j)}
[b_{vu}(u_{kl}-u_{ij})+b_{vv}(v_{kl}-v_{ij})],
\end{eqnarray}
where $i, j=1, \cdots, N$; $u$ and $v$ are the activator and inhibitor variables, respectively; $c$ is the coupling strength; and
$\epsilon>0$ is a small parameter characterizing a time scale separation, which we fix at $\epsilon=0.05$ in
this paper. Depending upon the threshold parameter $a$, each individual FHN unit exhibits either oscillatory
($|a|<1$) or excitable ($|a|>1$) behavior. We here fix it at $a=0.5$. Each element is coupled to its
neighbors within a finite range $r$ defined as
\begin{eqnarray}
\label{coupling}
B_r(i,j)&=&\{ (k,l): k\in [i-r, i+r], l\in [j-r, j+r], \nonumber \\
& & k\not=i, l\not=j \}.
\end{eqnarray}
Thus, the coupling region is a square of side $r$ around site $(i,j)$. From Eqs. (\ref{model}) and
(\ref{coupling}), we see that there are $(2r+1)^2-1$ coupling links from node $ij$ to the nodes in the
coupling region, and they constitute a starlike subnetwork. The entire network of system (\ref{model})
consists of all the $N^2$ subnetworks.
As in Refs. \cite{Omelchenko:2013,Hizanidis:2014}, we consider not only direct $u - u$ and
$v - v$ coupling, but also cross-coupling between variables $u$ and $v$. This feature is modeled by the
rotational coupling matrix
\begin{equation}\label{coupling-matrix}
B= \left(\begin{array}{cc}
b_{uu} & b_{uv}\\
b_{vu} & b_{vv}
\end{array}\right)
=\left(\begin{array}{cc}
\cos\phi & \sin\phi\\
-\sin\phi & \cos\phi
\end{array}\right),
\end{equation}
which depends on the coupling phase $\phi$. We let $\phi$ be slightly smaller than $\pi/2$; i.e.,
$\phi=\frac{\pi}{2}-0.1$.

To study the collective behaviors of system (\ref{model}) and investigate the existence of CSs, we introduce
two quantities. The first is the average phase velocity, which is defined as
\begin{equation}
\label{velocity}
\omega_{ij}=\frac{2\pi M_{ij}}{\Delta T}
\end{equation}
for the $ij$-th FHN unit, where $M_{ij}$ denotes the firing number of neuron $(i,j)$ in the time
period $\Delta T=t_2-t_1$. We here take $t_1=3000$ and $t_2=5000$. The second is the local order
parameter $R_{ij}$. To determine $R_{ij}$, we introduce the complex variable $z_{ij}=u_{ij}+iv_{ij}$. Then,
we choose a local region with radius $\delta$, i.e.,
\begin{equation}
\label{local}
Q_{\delta}(i,j)=\{ (k,l): k\in [i-\delta, i+\delta], l\in [j-\delta, j+\delta] \},
\end{equation}
which has an area of $(2\delta+1)^2$. When $\delta=0$, $Q_{\delta}(i,j)$ contains only the node $(i,j)$.
Comparing Eq. (\ref{local}) with Eq. (\ref{coupling}), we see that $Q_{\delta}(i,j)$ includes node $(i,j)$,
whereas $B_R(i,j)$ does not. Letting
\begin{equation}
\label{order}
Z_{ij}=\frac{1}{(2\delta+1)^2}\sum_{(k,l)\in Q_{\delta}(i,j)} z_{kl},
\end{equation}
the local order parameter $R_{ij}$ can be defined as
\begin{equation}
\label{order}
\langle Z_{ij}\rangle=\frac{1}{\delta t}\int_{t-\delta t}^tZ_{ij}(s)ds\equiv R_{ij}e^{i\psi_{ij}},
\end{equation}
with $\delta t=50$. $R_{ij}$ describes the local average amplitude of the $ij$-th FHN unit \cite{Matthews:1991,Cross:2006}.
A larger local order parameter $R_{ij}$ indicates that the $ij$-th unit belongs to the coherent part of the CS.

\section{Numerical simulations}

\begin{figure}
\epsfig{figure=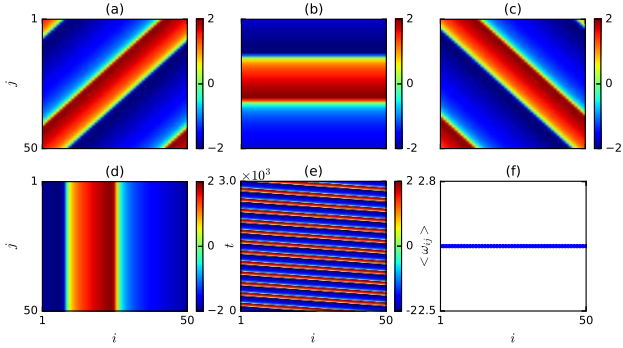,width=1.\linewidth} \caption{(color online.)
{\bf Traveling patterns}. (a)-(d): The distribution of the variable
$u_{ij}(t)$ is plotted on the $i-j$ plane for different parameters.
(a) Snapshot of the traveling wave along the diagonal with $c=0.1, r=7$ and $t=4700$.
(b) Snapshot of the traveling wave along the $j$ direction with $c=0.15, r=9$ and $t=4700$.
(c) Snapshot of the traveling wave along the clinodiagonal with $c=0.2, r=7$ and $t=5000$.
(d) Snapshot of the traveling wave along the $i$ direction with $c=0.2, r=10$ and $t=5000$.
(e) Traveling wave on the line with $j=25$ in (d).
(f) Distribution of the average phase velocity $\omega_{ij}$ in (e) with $j=25$.
}
\label{Fig:Traveling}
\end{figure}

{\bf Diversity of chimera-like patterns in the model of 2D arrays of FHN neurons with nonlocal coupling.}
We numerically simulate the above model of 2D arrays of FHN neurons with nonlocal coupling.
In the numerical simulations, we fix $N=50$ and let
the initial conditions be $u_{ij}^2(0)+v_{ij}^2(0)=4$. By changing the coupling strength $c$ and the coupling radius
$r$, we find that system (\ref{model}) can show different chimera-like patterns. For example,
Fig. \ref{Fig:Traveling}(a)--(d) show four typical patterns of traveling waves based on the distribution of the variable
$u_{ij}(t)$ on the $i-j$ plane, where (a) is a snapshot of the traveling wave along the diagonal with
$c=0.1, r=7$, and $t=4700$, (b) is a snapshot of the traveling wave along the $j$ direction with $c=0.15, r=9$, and
$t=4700$, (c) is a snapshot of the traveling wave along the clinodiagonal with $c=0.2, r=7$, and $t=5000$, and
(d) is a snapshot of the traveling wave along the $i$ direction with $c=0.2, r=10$, and $t=5000$. The traveling
behavior can be seen more clearly in Fig. \ref{Fig:Traveling}(e), which shows the behavior along the line $j=25$ in
Fig. \ref{Fig:Traveling}(d). We see that the behaviors in Fig. \ref{Fig:Traveling}(e) are periodic,
implying the same average phase velocity $\omega_{ij}$ on different $i$. This conclusion is confirmed in
Fig. \ref{Fig:Traveling}(f) by the distribution of the average phase velocity $\omega_{ij}$.

\begin{figure}
\epsfig{figure=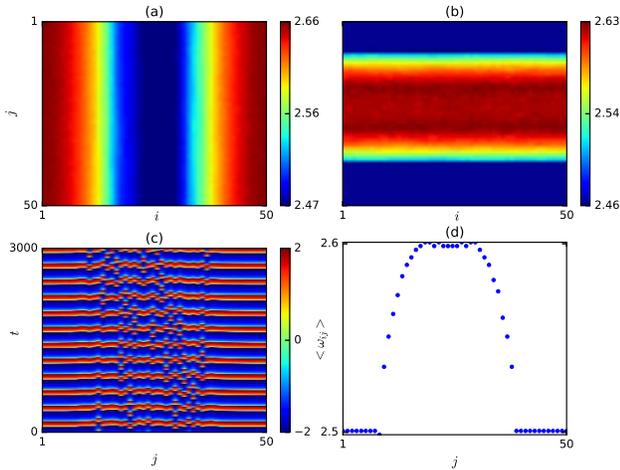,width=1.\linewidth}
\caption{(color online.) {\bf Stripes chimera states}. (a)-(b): The distribution of
$\omega_{ij}$ is plotted on the $i-j$ plane for different parameters.
(a) Case of $c=0.1$ and $r=16$.
(b) Case of $c=0.15$ and $r=19$.
(c) Evolution of the variable $u_{ij}(t)$ for the line with $i=25$ in (b).
(d) Distribution of the average phase velocity $\langle \omega_{ij}\rangle$ in (b) with $i=25$.}
\label{Fig:Stripes}
\end{figure}

\begin{figure}
\epsfig{figure=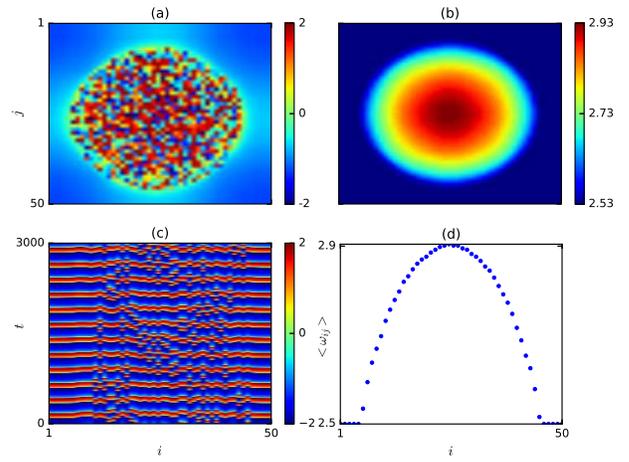,width=1.\linewidth}
\caption{(color online.) {\bf Circular spots chimera states}. (a) The distribution
of the variable $u_{ij}(t)$ is plotted on the $i-j$ plane for parameters $c=0.25$ and $r=16$ where
the snapshot of the circular spots is taken at $t=4500$.
(b) Distribution of $\omega_{ij}$ corresponding to (a).
(c) Evolution of the variable $u_{ij}(t)$ for the line with $j=25$ in (a).
(d) Distribution of the average phase velocity $\langle \omega_{ij}\rangle$ in (b) with  $j=25$.}
\label{Fig:spots}
\end{figure}

Fig. \ref{Fig:Stripes} shows two typical patterns of stripe CSs based on the distribution of $\omega_{ij}$
on the $i-j$ plane for (a) $c=0.1$ and $r=16$ and (b) $c=0.15$ and $r=19$.
To see the stripes clearly, Fig. \ref{Fig:Stripes}(c) shows the evolution of the variable $u_{ij}(t)$
for the line $i=25$ in Fig. \ref{Fig:Stripes}(b). We see that the central part is different from its two
sides. Fig. \ref{Fig:Stripes}(d) shows the distribution of the average phase velocity $\langle \omega_{ij}\rangle$
in Fig. \ref{Fig:Stripes}(b) with $i=25$. It is easy to see that the central part has a distribution of the
average phase velocity $\langle \omega_{ij}\rangle$, reflecting the stripes. Similarly,
Fig. \ref{Fig:spots} shows the patterns of a circular spot CS; see its caption for details.

\begin{figure}
\epsfig{figure=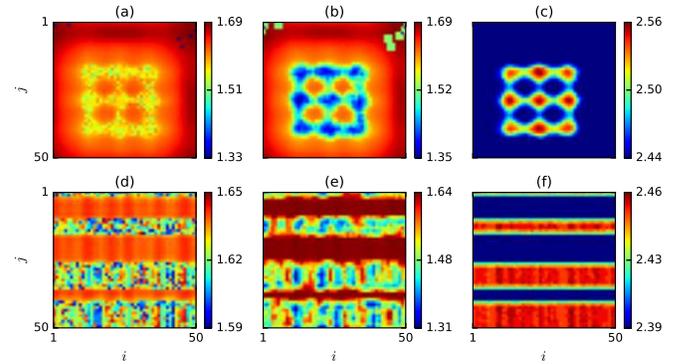,width=1.\linewidth} \caption{(color online.)
{\bf Typical patterns of gridding and multi-columns.}
(a)-(c) denote the case of gridding patterns with $c=0.3, r=12$ and $t=5000$,
where (a) represents the distribution of $R_{ij}$ for $\delta=0$; (b) the distribution of $R_{ij}$ for
$\delta=1$; and (c) the distribution of $\omega_{ij}$.
(d)-(f) denote the case of multi-columns patterns with $c=0.15, r=23$ and $t=5000$,
where (d) represents the distribution of $R_{ij}$ for $\delta=0$; (e) the distribution of $R_{ij}$ for
$\delta=1$; and (f) the distribution of $\omega_{ij}$.
}
\label{Fig:gridding}
\end{figure}

In addition to these typical CS patterns, we find, interestingly, that system (\ref{model}) can also show
special patterns such as gridding and multicolumn patterns, which have not been
observed in neural systems with both phase and amplitude before. Fig. \ref{Fig:gridding} shows typical
gridding patterns [(a)--(c)] and multicolumn patterns [(d)--(f)].

\begin{figure}
\epsfig{figure=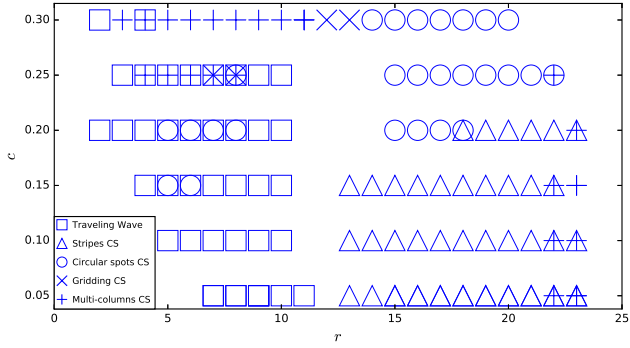,width=1.0\linewidth} \caption{(color online.)
{\bf Phase diagram of chimera-like patterns on the parameters $c-r$ plane.}
In this plane, the local regions with ``symbols" denote the existence of chimera-like patterns
and the empty regions imply no chimera-like patterns, where the ``squares'', ``triangles'',
``circles'', ``crosses'' and ``pluses'' represent the traveling, stripes, circular spots, gridding and
multi-columns patterns, respectively. }
\label{Fig:Phase-diagram}
\end{figure}

To clearly see the diversity of chimera-like patterns, Fig. \ref{Fig:Phase-diagram} shows the
phase diagram of chimera-like patterns on the
$c-r$ plane, where the squares, triangles, circles, crosses, and plus signs represent
the traveling, stripe, circular spot, gridding, and multicolumn patterns,
respectively. In this plane,
empty regions indicate the absence of CS patterns, and overlapping of different symbols indicates that multiple patterns may
appear at the same $c$ and $r$. More patterns may be found outside of this plane, indicating
that system (\ref{model}) can show a variety of chimera-like patterns.

\begin{figure}
\epsfig{figure=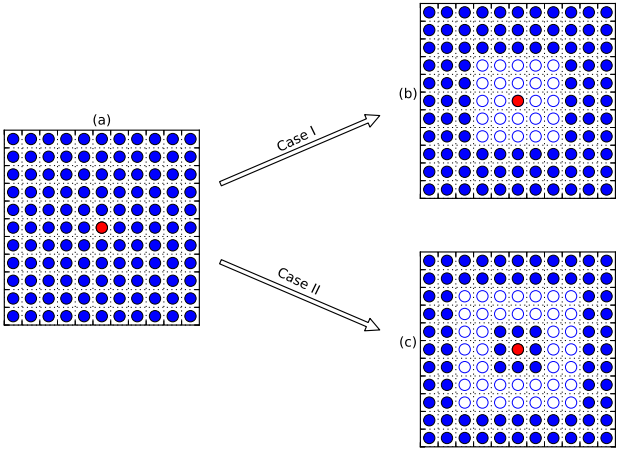,width=1.0\linewidth} \caption{(color online.)
{\bf Schematic illustration of removing part of the coupling links.} (a) The coupling scheme of node $ij$
before removing part of the coupling links where $r=5$ and the central ``red'' site denotes the chosen node-$ij$.
(b) The first approach to remove coupling links with $r_c=2$: remove all the links in the range $B_{r_c}(i,j)$ and
let the links in the range $B_{r}(i,j)-B_{r_c}(i,j)$ be remained. That is, the links between the ``red''
node and ``white'' nodes are removed, while the links between the ``red'' node and ``blue'' nodes are remained.
(c) The second approach to remove links: The parameters are chosen as $r=5$ and $m=2$, which gives $r_m=[2,3]$. Thus,
we remove all the middle links within the circles 2 and 3 and let the links on the first, fourth and fifth
circles be remained.   }
\label{Fig:scheme}
\end{figure}

{\bf Heterogeneous coupling obtained by removing some of the coupling links.}
In addition to changing the parameters $c$ and $r$, there may be other ways to change the coupling scheme
that can generate richer and more colorful chimera-like patterns. We here present two such approaches
by removing some of
the coupling links. The first approach is to remove all the links of node $ij$ within a radius $r_c$, where
$r_c\in [1, r-1]$. That is, we remove all the links of node $ij$ in the range
$B_{r_c}(i,j)=\{ (k,l): k\in [i-r_c, i+r_c], l\in [j-r_c, j+r_c], k\not=i, l\not=j \}$
and let the links in the range $B_{r}(i,j)-B_{r_c}(i,j)$ remain. Fig. \ref{Fig:scheme}(a) shows the
coupling scheme of node $ij$ before the coupling links are removed for $r=5$. Fig. \ref{Fig:scheme}(b) shows
a schematic illustration of the coupling links for the first approach with $r_c=2$. We
see that the number of coupling links remaining is $(2r+1)^2-(2r_c+1)^2=96$. Once this operation is performed,
the coupling matrix will be significantly changed.

\begin{figure}
\epsfig{figure=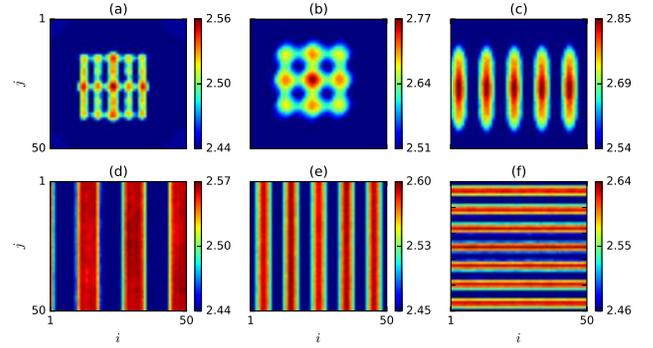,width=1.0\linewidth} \caption{(color online.)
{\bf Typical patterns of gridding and multi-columns for the first approach
to remove part of the coupling links.}
(a)-(f) represent the distributions of $\omega_{ij}$ for different parameters. (a)-(c) have the same initial
conditions and parameters with that in
Fig. \ref{Fig:gridding}(a)-(c) with $r=12$. The parameter $r_c$ is chosen as $r_c=3$ in (a), $r_c=6$ in (b), and
$r_c=7$ in (c). (d)-(f) have the same initial conditions and parameters with that in Fig. \ref{Fig:gridding}(d)-(f)
with $r=23$. The parameter $r_c$ is chosen as $r_c=15$ in (d), $r_c=17$ in (e), and $r_c=19$ in (f). }
\label{Fig:gridding1}
\end{figure}

We first let the initial conditions and parameters be the same as in Fig. \ref{Fig:gridding}(a)--(c) with
$r=12$. We find that the gridding pattern can still be observed when $r_c\in [1,6]$.
Figs. \ref{Fig:gridding1}(a) and
(b) show the distributions of $\omega_{ij}$ for two typical cases. Comparing them with Fig. \ref{Fig:gridding}(c),
we see that Fig. \ref{Fig:gridding1}(b) is similar to Fig. \ref{Fig:gridding}(c), but Fig. \ref{Fig:gridding1}(a)
has two more columns than Fig. \ref{Fig:gridding}(c), indicating that $r_c$ will influence the structure of
the gridding pattern. More interestingly, we find that when $r_c\in [7,9]$, the gridding
pattern is replaced by a pattern of
five oval region, as shown in Fig. \ref{Fig:gridding1}(c), indicating that the gridding pattern has been
transformed into a multicircle pattern. Then we let the initial conditions and
parameters be the same as in
Fig. \ref{Fig:gridding}(d)--(f) with $r=23$. We find
that we can still observe the pattern of three columns when $r_c\in [1,2]U[9,15]$. Fig. \ref{Fig:gridding1}(d)
shows such a typical result with $r_c=15$. However, we observe a pattern of five columns when
$r_c\in [3]U[5,8]U[16,18]$ and a pattern of seven columns when $r_c\in [4]U[19,20]$. Figs. \ref{Fig:gridding1}(e)
and (f) show typical results for $r_c=17$ and $19$, respectively.

We observed similar results for other cases such as the traveling wave, stripes, and circular spots
and found that their patterns differ greatly from those in
Figs. \ref{Fig:Traveling}, \ref{Fig:Stripes}, and \ref{Fig:spots}, respectively, confirming the diversity of chimera-like patterns.

The second approach is to remove all the coupling links in the middle of a specific range but keep at least the nearest
and farthest coupling links. Specifically, we let $m$ be the number of removed circles and let $m\in [1,r-2]$, which
determines how many circles of coupling links will be removed. In this sense, the coupling links within the radius
$r_m=[int(r/2)+1-int(m/2), int(r/2)+1+int((m-1)/2))]$ will be removed. Fig. \ref{Fig:scheme}(c) illustrates
schematically the coupling links for the second approach with $r=5$ and $m=2$. We
see that the remaining links are in two separate regions, and the removed links are in the range of $r_m=[2,3]$,
indicating that all the links in circles 2 and 3 are removed. This approach makes the coupling matrix more
heterogeneous, in contrast to the first approach.

\begin{figure}
\epsfig{figure=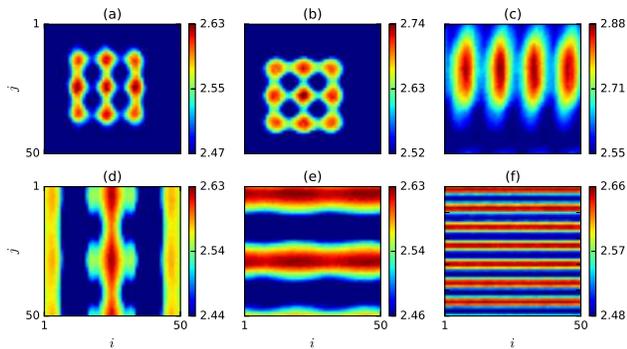,width=1.0\linewidth} \caption{(color online.)
{\bf Typical patterns of gridding and multi-columns for the second approach to
remove coupling links.} (a)-(f) represent the distributions of $\omega_{ij}$ for different parameters.
(a)-(c) have the same initial conditions and parameters with that in Fig. \ref{Fig:gridding}(a)-(c) with $r=12$.
The parameter $m$ is chosen as $m=1$ in (a), $m=2$ in (b), and $m=8$ in (c).
(d)-(f) have the same initial conditions and parameters with that in Fig. \ref{Fig:gridding}(d)-(f) with $r=23$.
The parameter $m$ is chosen as $m=14$ in (d), $m=15$ in (e), and $m=18$ in (f). }
\label{Fig:gridding2}
\end{figure}

To see how changing the coupling matrix influences the chimera-like patterns, we first let the initial
conditions and parameters be the same as those in Fig. \ref{Fig:gridding}(a)--(c) with $r=12$. We use the second
approach to remove coupling links and find that the gridding pattern can still be observed
when $m\le 2$. Figs. \ref{Fig:gridding2}(a) and (b) show the distributions of $\omega_{ij}$ for $m=1$ and $2$,
respectively. We see that they are similar to those in Fig. \ref{Fig:gridding}(c). However,
we find that when $m=8$, the gridding pattern is replaced by a pattern of four oval
region, as shown in Fig. \ref{Fig:gridding2}(c). Then we let the initial conditions and parameters be the same as
those in Fig. \ref{Fig:gridding}(d)--(f) with $r=23$. We find that we can still observe the pattern of three columns when
$m\le 8$. However, we observe a pattern of two columns when $m\in [15,17]$; the coexistence of a circular spot and a
stripe when $m=9, 11, 13, 14$; two stripes when $m=10$ and $12$; and a pattern of seven columns when $m=18$.
Figs. \ref{Fig:gridding2}(d)--(f) show the typical results for $m=14, 15$, and $18$, respectively. We can also compare
each panel in Fig. \ref{Fig:gridding2} with the corresponding panel in Fig. \ref{Fig:gridding1}. We
easily see that the panels in the two figures are different, indicating the diversity of the chimera-like patterns.

We observed similar results for other cases such as the traveling wave, stripes, and circular spots
using this approach, again confirming the diversity of patterns.

\section{Discussions and conclusion}

The networked neural model in Eq. (\ref{model}) is in fact not a phase model. Compared with phase models,
it contains more information, including both the aspect from phase and that from
amplitude. In this sense, much more information will be transmitted by the coupling links than
in a purely phase model, and thus the system will be able to show a richer variety of behavior, which is the main reason for the diversity of chimera-like patterns in a single networked neural system.

A characteristic feature of Eq. (\ref{model}) is its nonlocal coupling, which is fundamentally different from
both global coupling and nearest-neighbor coupling. Global coupling generally induces global
synchronization that is robust to slight changes in the network structure. In contrast, nearest-neighbor coupling does not easily result in global synchronization unless a sufficiently large coupling strength is
used. Nonlocal coupling falls between these two extremes. Thus, it cannot cause global synchronization
but can only sustain local synchronization, which guarantees the existence of different chimera-like patterns.

More importantly, we present a new way of changing the structure of coupling links, which is the guarantee of
the diversity of the chimera-like patterns. Unlike the use of homogeneously nonlocal coupling in a region of radius $r$,
removing some of the coupling links yields heterogeneously nonlocal coupling and provides more
possible coupling structures, greatly increasing the diversity of the chimera-like patterns. On the other hand, this
idea can be easily extended to different schemes. Notice that in both cases I and II in Fig. \ref{Fig:scheme},
the number of coupling links remains identical for each node. Therefore, an easy way to extend them is to
design irregular coupling schemes in which the number of coupling links may be different for different nodes.
For example, we may randomly remove some of the coupling links. In particular, we may also remove the coupling links
on purpose so that the final coupling networks can meet desired criteria, such as scale-free networks or other complex
networks. In this way, we may expect new chimera-like patterns.

In conclusion, we observed diverse chimera-like patterns in 2D arrays of FHN neurons with nonlocal coupling,
which were previously observed only in 1D arrays or 2D arrays in phase models. The numbers of both coupling schemes and the resulting
chimera-like patterns were vastly increased by removing some of the coupling links.

This work was partially supported by the NNSF of China under Grant
Nos. 11135001 and 11375066, 973 Program under Grant No. 2013CB834100, the Scientific Research Starting Foundation of Tongren University under Grant No. TS1118 and the NSF of Guizhou Province Education Department under Grant No. KY[2014]316.

\end{document}